\documentclass[12pt, a4paper,preprint,showpacs,preprintnumbers]{article}
\usepackage{a4wide,amssymb,amsmath,amscd}
\usepackage{color}

\begin{document}

\topmargin -2pt


\headheight 0pt

\topskip 0mm \addtolength{\baselineskip}{0.20\baselineskip}

\vspace{5mm}

\begin{center}
{\Large \bf Bound of Noncommutativity Parameter \\
Based on Black Hole Entropy} \\

\vspace{10mm}

{\sc Wontae Kim}${}^{\dag,}$\footnote{wtkim@sogang.ac.kr} and {\sc Daeho Lee}${}^{\ddag, }$\footnote{dhleep@sogang.ac.kr} \\

\vspace{1mm}

${}^{\dag}${\it Department of Physics, Sogang University, Seoul,
121-742, Korea, Center for Quantum Spacetime, Sogang University, Seoul 121-742, Korea,
and School of Physics, Korea Institute for Advanced Study, Seoul 130-722, Korea}\\
${}^{\ddag}${\it Basic Science Research Institute, Sogang University, Seoul 121-742, Korea}\\

\vspace{10mm}
{\bf ABSTRACT}
\end{center}

\noindent
\noindent
We study the bound of the noncommutativity parameter in the
noncommutative Schwarzschild black hole which is a solution of the
noncommutative $ISO(3,1)$ Poincar$\acute{e}$ gauge group.
The statistical entropy satisfying the area law in the brick wall method
yields a cutoff relation which depends on the noncommutativity parameter.
Requiring both the cutoff parameter and the noncommutativity parameter
to be real, the noncommutativity parameter can be shown to be bounded
as $\Theta > 8.4\times 10^{-2}l_{p}$.
\vfill

\thispagestyle{empty}

\newpage
%

%
Since short distance behaviors have been
extensively studied, it has been claimed
that the description of spacetime as a discontinuous manifold may be
plausible. In this respect, it seems to be natural to consider
spacetimes related with noncommutativity
in which the coordinates become noncommutative.
Its notion has been popular
in the context of the string theory and
intrinsically connected with gravity \cite{cds,sw99,dfr95}.
Explicitly, canonical commutation relations for noncommutative
spacetime are assumed to be
\begin{equation}
\label{moyal_nc}
[\hat{x}^\alpha,\hat{x}^\beta]=i \Theta^{\alpha\beta},
\end{equation}
where $\Theta^{\alpha\beta}$ are
anti-symmetric tensors. It has been well known that a theory on
noncommutative spacetime along with the commutation relation
\eqref{moyal_nc} is equivalent to another theory on commutative
spacetime in which a product of any two functions on the original
noncommutative spacetime is replaced with the Moyal
star$(\star)$-product \cite{groen,moyal}:
\begin{equation}
\label{moyalprd}
(f\star g)(x)\equiv \left.\exp\left[\frac{i}{2}\Theta^{\alpha\beta}\frac{\partial}
{\partial x^{\alpha} }\frac{\partial}{\partial y^{\beta} }\right] f(x)g(y)\right|_{x=y} .
\end{equation}
Such a deformation in a gravity side can be constructed based
on gauging the noncommutative $SO(4,1)$ de Sitter group  and
the Seiberg-Witten map
with subsequent contraction to $ISO(3,1)$ Poincar$\acute{e}$
gauge group \cite{cham01,ms06,bms07}.

Various spherically symmetric black hole
solutions and cosmological solutions in the commutative spacetime
are promoted to the noncommutative ones
through the Seiberg-Witten map \cite{ctz08,ms08,ctsz08,lmmp02}.
In particular, the metric of the noncommutative Schwarzschild black hole
up to the second order of the noncommutativity parameter is given by \cite{ctz08}
\begin{equation}
\label{ncmetric}
ds^2=g_{tt} dt^2+g_{rr}dr^2+g_{\theta\theta}d\theta^2+g_{\phi\phi}d\phi^2,
\end{equation}
where
\begin{eqnarray}
g_{tt} &=&-\left[ \left(1-\frac{2GM}{r}\right)+\frac{GM(4r-11GM)}{4r^4}\Theta^2\right] \nonumber \\
g_{rr} &=& \left(1-\frac{2GM}{r}\right)^{-1}-\frac{GM(2r-3GM)}{4r^2(r-2GM)^2}\Theta^2
\nonumber \\
g_{\theta\theta} &=& r^2+\frac{r^2-17GM r+34G^2 M^2}{16r(r-2GM)}\Theta^2
\nonumber \\
g_{\phi\phi} &=& r^2\sin^2\theta+\frac{(r^2+2GM
  r-4G^2M^2)\cos^2\theta-4GM(r-GM)}{16r(r-2GM)}\Theta^2, \nonumber
\end{eqnarray}
where the noncommutativity parameter $\Theta$ is defined by the commutation relations
\begin{equation}
\label{commrel_polar}
[\hat{r},\hat{\theta}]=i\Theta,~~\texttt{others}=0.
\end{equation}
By the way, the commutation relation (\ref{commrel_polar}) is different from
the usual cartesian one (\ref{moyal_nc}) since it corresponds
to $\Theta^{ij}=r\Theta \epsilon^{ij}$. However, the Moyal
star$(\star)$-product (\ref{moyalprd}) can be consistently
defined in the spherical coordinates with a constant
$\Theta^{r\theta}(\equiv \Theta)$. Therefore, the noncommutativity
parameter $\Theta$ carries a length dimension.
Actually, the metric solution \eqref{ncmetric} was derived based on
the assumption of the spherical representation rather than the
cartesian coordinates \cite{ctz08}.

Here $G=l_{p}^2$ and $M$  are the Newton's constant and the total mass of the black hole,
respectively. Note that the metric is not spherically symmetric unlike the case of the commutative
Schwarzschild black hole in general relativity.
It has coordinate singularities, such as apparent and Killing
horizons, $\hat{r}_{H}=2GM$ and $\tilde{r}_{H}=2GM \left(1+ \frac{3\Theta^2} {64G^2M^2}\right)$, respectively.
Here, the apparent and Killing horizons are defined by $g^{rr}=g_{rr}^{-1}=0$ and $g_{tt}=0$ at each horizon in the usual context, respectively.

Note that there have been some efforts to determine the bounds of
the noncommutative parameter \cite{saha07,bbdp09,nss06,alavi09}. In Refs. \cite{nss06,alavi09},
it has been shown that the bound of the noncommutativity parameter
appears at the order of $10^{-1} l_p $ by taking into account
the significant back reaction of the thermal
temperature should be the same order of magnitude as the total mass in
the Gaussian extension of the point source induced by the
noncommutativity.
In this paper,
we would like to study the bound of the noncommutativity parameter in
such a different way that the entropy of the noncommutative Schwarzschild black hole
should be satisfied with the well-known area law.
The entropy-area relationship in the brick-wall method
\cite{thooft,jy99,kop01,wy04,kins06,rlhy08,ksy08,ely09}
yields a relation between the brick
wall cutoff and the noncommutativity parameter. Remarkably,
the real condition of the relation will give the bound of the
noncommutativity parameter, which is the same order of that in Refs. \cite{nss06,alavi09}.

Now, let us consider a scalar field confined in a box near the Killing
horizon of the black hole. Along with the metric (\ref{ncmetric}),
the equation of motion for the scalar field on this black hole
background is,
\begin{eqnarray}
\label{kle}
\frac{1}{\sqrt{-g}}\partial_{\mu}(\sqrt{-g}g^{\mu\nu}\partial_{\nu}\Phi)-\mu^2 \Phi=0
\end{eqnarray}
with the boundary condition,
 $\Phi(x)=0~~\texttt{for}~~ r\leq \tilde{r}_{H}+\epsilon~\texttt{and}~~ r\geq L$
 where $\tilde{r}_{H},  \tilde{r}_{H}+\epsilon$ and $L$ are the positions
of the Killing horizon, the inner and outer walls of a spherical
box, respectively. We assumed that a quantum gas is in a thermal
equilibrium state at the temperature $T=\beta^{-1}$.

In the WKB approximation with $\Phi=e^{-iEt+iS(r,\theta,\phi)}$,
the field equation \eqref{kle} yields the constraint
\begin{equation}
\frac{p_{r}^2}{g_{rr}}+\frac{p_{\theta}^2}{g_{\theta\theta}}+\frac{p_{\phi}^2}{g_{\phi\phi}}
=((-g^{tt})E^2-\mu^2),
\end{equation}
where the momenta $p_{i}$'s are defined by $p_{r}=\partial S/\partial r, p_{\theta}=\partial S/\partial \theta, p_{\phi}=\partial S/\partial \phi$.
Then, according to the semiclassical quantization rule, the number of quantum states $n(E)$
with energy not exceeding $E$ can be written as
\begin{eqnarray}
n(E) &=& \frac{1}{(2\pi)^3} \int dr d\theta d\phi dp_r dp_{\theta}dp_{\phi}
\nonumber \\
&=& \frac{1}{6\pi^2} \int drd\theta d\phi \sqrt{g_{rr} g_{\theta\theta} g_{\phi\phi}}
[(-g^{tt}E^2-\mu^2)]^{3/2}.
\end{eqnarray}
Then, the free energy is found to be
\begin{eqnarray}
\label{free_energy}
F &=& -\int dE \frac{n(E)}{e^{\beta E}-1}
\nonumber \\
&=& -\frac{\pi^2}{90 \beta^4} \int_{\tilde{r}_{H}+\epsilon}^{L} dr (-g^{tt})^{3/2} g_{rr}^{1/2} \int d\theta d\phi (g_{\theta\theta} g_{\phi\phi})^{1/2},
\end{eqnarray}
where the mass of the scalar field is set to be zero for simplicity.
In the near horizon, it reads
\begin{eqnarray}
\label{refree_energy}
F &\simeq & -\frac{\pi^2}{90 \beta^4} \left[\int_{\tilde{r}_{H}+\epsilon}^{L} dr (-g^{tt})^{3/2} g_{rr}^{1/2}\right]
\left[\int d\theta d\phi (g_{\theta\theta} g_{\phi\phi})^{1/2}\right]_{\tilde{r}_{H}}
\nonumber \\
&=&  -\frac{\pi^2 A_{H}}{90 \beta^4} \int_{\tilde{r}_{H}+\epsilon}^{L} dr (-g^{tt})^{3/2} g_{rr}^{1/2},
\end{eqnarray}
where $A_{H}$ is the surface area of the noncommutative Schwarzschild black hole on the Killing horizon $\tilde{r}_{H}$.
The degrees of freedom of the scalar field are assumed to be dominated in the
vicinity of the horizon, so that one can use the near horizon approximation.
To calculate the radial integration of the
free energy, it gives
\begin{eqnarray}\label{appfree_energy}
F \approx -\frac{8\pi^2G^2M^2(2GM)^{1/2} A_{H}}{45\beta^4\Theta^2}\sqrt{\frac{8GM \epsilon+\Theta^2}{\epsilon}},
\end{eqnarray}
where we ignored the infra-red contribution coming from the outer boundary $L$
in the radial integration \cite{thooft}.
Now, we need Hawking temperature $\tilde{T}_{H}$, which is defined
by the Killing vector $K=\partial_{t} $ as
\begin{equation}
\tilde{T}_{H}=\frac{\tilde{\kappa}_{H}}{2\pi}=
\frac{1}{8\pi GM}\left(1+\frac{7\Theta^2}{32G^2M^2}\right)
= \tilde{\beta}_{H}^{-1},
\end{equation}
where $\tilde{\kappa}_{H}$ is the surface gravity at the Killing
horizon. From now on, we will assume $\Theta^2/G^2M^2 \ll 1$ for
later convenience, which means that the black hole mass is much larger than $\Theta /l_p^2$.
Using the thermodynamic relation,
$S=\beta^{2} \partial F/\partial\beta|_{\beta=\tilde{\beta}_{H}}$,
the statistical entropy from \eqref{appfree_energy} simply becomes
\begin{eqnarray}
\label{ns_entropy}
S_{NS} &=& \frac{A_{H}}{180\pi \Theta^2}
  \sqrt{\frac{\bar{\epsilon}^2+2\bar{\epsilon}\Theta+\Theta^2}
  {\bar{\epsilon}^2+2\bar{\epsilon}\Theta}},
\end{eqnarray}
where the cutoff parameter $\epsilon$ was replaced with a proper
length,  $\bar{\epsilon} \approx  \sqrt{8GM\epsilon+\Theta^2}-\Theta$.
The entropy is different from that of the Schwarzschild black hole
in that it depends on the noncommutativity parameter.

It is interesting to note that
the classical metric solution \eqref{ncmetric} has a well-defined commutative
limit for the vanishing noncommutativity parameter, while the
resulting entropy does not. The reason comes from the radial integration of 
the free energy \eqref{refree_energy}. For instance, just like the integral of a
function $ e^{\alpha r}$ with a parameter $\alpha$, the vanishing
limit of $\alpha=0$ is well-defined, however, this is not the case after integration.
Similar calculations appear in performing the radial integration in
Eq. \eqref{refree_energy}, so that the resulting entropy does not have
a smooth vanishing limit.

Let us assume that the entropy \eqref{ns_entropy} is satisfied with the well-known area
law even in this noncommutative black hole,
\begin{equation}
S_{NS} = A_{H}/4l_{p}^2,
\end{equation}
which gives the following relation,
\begin{equation}
\label{gamma}
\bar{\epsilon} =\Theta \left(\sqrt{\frac{\gamma}{\gamma-1}}-1\right),
\end{equation}
where $\gamma=(\sqrt{45\pi} \Theta/ \l_{p})^4 $.
From the additional condition of $\gamma > 1$ to make the
noncommutativity parameter to be real,
it is eventually bounded as
\begin{equation}
\label{Theta}
\Theta > \frac{1}{\sqrt{45\pi}}l_{p} \sim 8.4\times 10^{-2}l_{p}.
\end{equation}
In particular, there is a critical case of $\gamma =4/3$ where the
noncommutative parameter is the same with the brick wall cutoff.

Now, let us mention the thermodynamic stability of the
noncommutative Schwarzschild black hole. Using the thermodynamic
relations, one calculate the internal energy of the black
hole system as
$U=\left.\frac{\partial}{\partial \beta}(\beta F)\right|_{\tilde{\beta}_{H}}
= \frac{3A_{H}}{128\pi G^2 M}$.
Then, the heat capacity can be evaluated
as
$C_{v}=\frac{\partial U}{\partial \tilde{T}_{H}}=
-\frac{3\pi GM^2}{2}I(\theta)
\left(1-\frac{3\Theta^2}{4G^2M^2}\right)$,
where the function $I(\theta)$ is explicitly given by
$I(\theta)=\int_{0}^{\pi} d\theta \frac{\sin^2\theta}
{\sqrt{\sin^2\theta
+\frac{\Theta^2}{128G^2M^2}(1-19\sin^2\theta)}}$,
which is well-defined positive definite function for
$\Theta^2/ G^2 M^2 \ll 1 $.
As a result, it is negative and thus our noncommutative black hole
is thermodynamically unstable as like the commutative case \cite{thooft}.

We have shown that the noncommutative parameter $\Theta$ can be
bounded by requiring the
standard area law of the entropy in the noncommutative
Schwarzschild black hole using the brick-wall method. Furthermore,
the heat capacity shows that the black hole is still
unstable even in spite of the noncommutativity.
It is interesting to note that the order of the bound \eqref{Theta}
is coincident with  $\sqrt{\theta} \gtrsim 10^{-1}l_{p}$ in
Refs. \cite{nss06,alavi09}, where $\theta$ has the squared length dimension.



\section*{Acknowledgments}
W. KIM was supported by the Special Research
Grant of Sogang University, 200911044, and
D. LEE was supported by the National Research Foundation
of Korea Grant funded by the Korean Government, NRF-2009-351-C00111.


\end{document}